\documentclass{emulateapj}
\usepackage[english]{babel}     
\usepackage{enumitem}
\usepackage{amssymb,amsmath}    
\usepackage{graphicx}                   
\usepackage{braket}             
\usepackage{hyperref}
\usepackage{natbib}
\usepackage{bm}

\slugcomment{First submitted to ApJ: 9 July 2015}

\shorttitle{The compact `ES' galaxy NGC 1271}
\shortauthors{Graham, Ciambur \& Savorgnan} 

\begin{document}

\title{Disky elliptical galaxies
and the allegedly over-massive black hole in the compact massive `ES' galaxy NGC 1271}

\author{Alister W.\ Graham, 
Bogdan C.\ Ciambur and 
Giulia A.D.\ Savorgnan}
\affil{Centre for Astrophysics and Supercomputing, Swinburne University of
  Technology, Victoria 3122, Australia.}
\email{AGraham@astro.swin.edu.au}

\begin{abstract}

While spiral and lenticular galaxies have {\it large-scale} disks extending
beyond their bulges, and most local early-type galaxies with 
$10^{10}<M_*/M_{\odot}<2\times10^{11}$ contain a disk (e.g., ATLAS$^{\rm 3D}$),
the early-type galaxies do possess a range of disk sizes.  The edge-on, {\it
  intermediate-scale} disk in the `disky elliptical' galaxy NGC~1271 has led to
some uncertainty as to what is its spheroidal component.  Walsh et~al.\ 
reported a directly measured black hole mass of
$(3.0^{+1.0}_{-1.1})\times10^9~M_{\odot}$ 
for this galaxy; which they remarked was an order of magnitude
greater than what they expected based on their derivation of the host 
spheroid's luminosity.  Our near-infrared image analysis supports a small 
embedded disk within a massive spheroidal component with 
$M_{sph,*}=(0.9\pm0.2)\times 10^{11}~M_{\odot}$ 
(using $M_*/L_H = 1.4^{+0.13}_{-0.11}$ from Walsh et~al.).  This places NGC~1271 
just 1.6-sigma above the near-linear $M_{\rm bh}$--$M_{sph,*}$ relation
for early-type galaxies.  Therefore, past speculation that there may be a
systematic difference in the black hole scaling relations between compact
massive early-type galaxies with intermediate-scale disks, i.e.\ ES galaxies
such as NGC~1271, and early-type galaxies with either no substantial disk (E)
or a large-scale disk (S0) is not strongly supported by NGC~1271.  
We additionally (i) show how ES 
galaxies fit naturally in the (`bulge'-to-total)-(morphological type)
diagram, while noting a complication with recent revisions to the Hubble-Jeans 
tuning-fork diagram, (ii) 
caution about claims of over-massive black holes in other ES galaxies
if incorrectly modelled as S0 galaxies, 
and (iii) reveal that the compact massive spheroid in NGC~1271 has properties
similar to bright bulges in other galaxies which have grown larger-scale disks.

\end{abstract}


\section{Introduction}

Over the past decade, there has been an increasing interest in the fate of the
compact, massive galaxies seen at $z\sim2\pm0.5$ (e.g.\ Daddi et al.\ 2005;
Toft et al.\ 2007; Trujillo et al.\ 2007; Damjanov et al.\ 2009, 2014; 
Weinzirl et al.\ 2011; Prieto et al.\ 2013; 
van Dokkum et al.\ 2008, 2015; Carollo et al.\ 2013; 
Zahid et al.\ 2016; Andreon et al.\ 2016).  
In mid-2011 Graham pointed 
out that these galaxies have the same physical properties as the spheroidal
component of some local lenticular galaxies (Graham 2013). Dullo \& Graham
(2013) further advocated that disks may have formed in and around these
high-$z$ mass concentrations to build the local lenticular
galaxies.  Nearly two dozen $z\approx0$ galaxies with massive compact spheroidal
components have since been presented in Graham, Dullo \& Savorgnan (2015, hereafter
GDS15; see also Valentinuzzi et al.\ 2010a,b; Poggianti et al.\ 2013a,b; Saulder
et al.\ 2015; Carollo et al.\ 2016; de la Rosa et al.\ 2016).  This sample included early-type
galaxies in which the disk had not grown into a large-scale disk that
dominates the light at large radii but was instead an intermediate-scale
disk\footnote{Although we use the term {\it intermediate-scale} disk to
  distinguish it from both {\it large-scale} disks which dominate the light at
  large radii, and {\it nuclear} disks which are typically tens to a few hundred
  parsec in size (e.g.\ Balcells et al.\ 2007), we point out that a continuum
  of disk sizes exists.} embedded within the spheroidal component of the
galaxy\footnote{Intermediate-scale disks are not just confined to compact
  massive galaxies; the dwarf early-type galaxy LEDA~074886 also contains an
  embedded, nearly edge-on, disk (Graham et al.\ 2012).}.  Liller (1966)
discovered such galaxies, bridging the E and S0 morphological types (Hubble
1936), and she designated them {\it ES} galaxies.  They have subsequently been
referred to as ``disk ellipticals'' (Nieto et al.\ 1988) and ``disky
ellipticals'' (Simien \& Michard 1990; Michard \& Marchal 1993; Andreon et
al.\ 1996).  Obviously, when a disk grows 
sufficiently large, the {\it total} galaxy is no longer compact even though
the spheroidal component still is.  Such disk growth would therefore result in
an apparent decrease in the number density of compact {\it galaxies} with
decreasing redshift but no reduction to the actual number of the compact {\it
  spheroids} themselves.

NGC~1277 was one such ES galaxy noted in GDS15 --- after van den Bosch et
al.\ 2012 identified the galaxy as compact and massive\footnote{Graham et
  al.\ (2016) have identified its spheroid component as compact and massive.}
--- as was NGC~1332 (Savorgnan \& Graham 2016b), NGC~5493 (Krajnovi\'c et
al.\ 2013) and NGC 5845 (Jiang et al.\ 2012).  NGC~1271 is another nearby,
compact early-type galaxy (Brunzendorf \& Meusinger 1999) that contains a
nearly edge-on, intermediate-scale disk.  It was first recognized as
having an unusually small half-light radius for its luminosity in the catalog
of Strom \& Strom (1978), with such galaxies labelled as `compact galaxies'
as far back as Zwicky \& Kowal (1968) and Zwicky \& Zwicky (1971).

Galaxies with intermediate-scale disks can be more akin to the spheroidal
component of lenticular galaxies that acquired large-scale disks than they are
to complete lenticular galaxies.  As noted above, although such
intermediate-scale disks are not unusual, they may be somewhat unfamiliar to
readers more versed with the bulge/disk structure of spiral and lenticular
galaxies.  The spheroidal stellar component of spiral galaxies presents itself
as a central bulge encased within a disk: that is, the spheroid does not
dominate the light at large radii --- as noted by the warning in Figure~7 from
Graham 2001).  However, early-type galaxies {\it can} have spheroidal
components which dominate the galaxy light at both inner and outer radii.
This situation can arise when a disk has not grown large enough to dominate
the light at large radii.  Galaxies in clusters may well be particularly prone
to this situation due to ram pressure stripping (Gunn \& Gott 1972) removing
their disk-building gas supply, and ``Strangulation'' (Larson et al.\ 1980;
Bekki et al.\ 2002) --- which also operates in galaxy groups (Kawata \&
Mulchaey 2008) --- cutting off their gas supply.  This may contribute to the
compact massive galaxies found in clusters by the WIde-field Nearby
Galaxy-cluster Survey (WINGS) collaboration (e.g.\ Valentinuzzi et al.\ 2010a;
see also Nantais et al.\ 2013). 

Careful bulge/disk decompositions are required if one is to avoid comparing
`apples and oranges' and subsequently misinterpreting systematic offsets in
various parameter scaling diagrams.  For example, it is not appropriate to
consider such small disks as the `bulge' component of a galaxy, nor to
consider only the inner portion of the spheroidal component (where its light
produces a central bulge in the radial light profile above the
intermediate-scale disk) as the complete spheroidal component of a galaxy, nor
to confuse the inner portion of the spheroid as an encased bulge surrounded by
a disk plus a separate envelope that is actually the outer portion of the
spheroid.

Considering the `bulge' component of a galaxy to be the smallest-sized
component in their multiple S\'ersic model descriptions of NGC~1271, Walsh et
al.\ (2015; hereafter W2015) derived a bulge mass for NGC~1271 of
$5.4\times10^{10}~M_{\odot}$.  This was the midpoint of their two estimates
($8.7\times10^{10}~M_{\odot}$ and $2.3\times10^{10}~M_{\odot}$) obtained by
fitting 2 and then 3 S\'ersic components to the galaxy image.  Their
uncertainty as to the actual spheroidal component of this galaxy led them to
adopt the average mass from the smallest-sized component of both
decompositions, which they then used in the near-linear $M_{\rm bh}$--$M_{\rm
  sph,*}$ relation from Kormendy \& Ho (2013) to predict a central black hole
mass that was an order of magnitude lower than their directly measured black
hole mass of $3.0^{+1.0}_{-1.1}\times 10^9~M_{\odot}$.  That is, they reported
the detection of an ``over-massive'' black hole in NGC~1271 relative to its
bulge (their Figure~10).  From this, they concluded that there could be
systematic differences in the black hole scaling relations between compact
massive galaxies (i.e.\ spheroids with intermediate-scale disks such as
NGC~1271 and NGC~1277) and the other spheroids which have been used to
construct the (black hole)--(host spheroid) mass scaling relation, i.e.\ the
spheroidal component of: galaxies with larger-scale disks; normal elliptical
galaxies; and brightest cluster galaxies.  W2015 went on to remark that a
difference in the black hole scaling relations implies a different growth
path, i.e.\ evolution, for black holes in such galaxies (see also
Ferr\'e-Mateu et al.\ 2015).

Given that the above interpretation effectively rests on the mass of the black
hole predicted from the stellar mass of the host spheroid, in
Section~\ref{Sec_Intro} we perform a spheroid/disk decomposition of NGC~1271
to independently determine the mass of its spheroidal component, and then
predict the mass of its central black hole using the latest $M_{\rm
  bh}$--$M_{\rm sph,*}$ relation\footnote{Graham (2016) provides a detailed
  review of the development of the $M_{\rm bh}$--$M_{\rm sph}$ relation, from
  Dressler (1989) and Yee (1992) up until the bent relation in Graham \& Scott
  (2015; see also the models by Lu \& Mo 2015 and Fontanot, Monaco \& Shankar
  2015).  In addition to describing the $M_{\rm bh}$--$M_{\rm sph,*}$
  relation, Graham (2016) contains an extensive bibliography to explain how
  and why we came to believe in (super-massive) black holes.}.  In addition to
the surface brightness profile, we use both the ellipticity profile and the
``disky / boxy'' $B_4$ profile --- which quantifies the deviations of the
isophotes from ellipses (Carter 1978, 1979, 1987; Jedrzejewski 1987; Bender \&
M\"ollenhoff 1987; Ebneter et al.\ 1988; Bender 1988; Bijaoui et al.\ 1989)
--- to reveal the radial extent of the disk in NGC~1271.  We do so using the
routine described in Ciambur (2015), which improves upon the prescription used
in the popular {\sc iraf} task {\sc ELLIPSE} (Jedrzejewski 1987).  We show
that the disk resides wholly within the spheroidal component of NGC~1271 which
dominates the light at large radii.  Our analysis results in a larger spheroid
mass and thus a lower (black hole)-to-spheroid mass ratio than was recently
highlighted by W2015.  In Section~\ref{Sec_Disc} we discuss this result,
placing NGC~1271 in the $M_{\rm bh}$--$M_{\rm sph,*}$ diagram, and we
additionally reveal that the compact spheroidal component of NGC~1271 is also
not an outlier in either the mass-size or the mass-density diagram.  We go on
to discuss the evolution, or lack thereof, of compact massive spheroids, and
also reveal how ES galaxies relate to other galaxies in a familiar
classification diagram involving the spheroid-to-disk flux ratio and galaxy
morphological type.  Our main conclusions are briefly presented in
Section~\ref{Sec_Conc}.

Following W2015, we adopt their Perseus galaxy cluster (Abell
426) distance 
of 80 Mpc to NGC~1271.  Together with their use of $H_0 = 70.5$ km s$^{-1}$
Mpc$^{-1}$, $\Omega_m = 0.27$ and $\Omega_{\Lambda} = 0.73$, this gives a
Hubble expansion redshift $z=0.0188$ and a scale of 379 parsec per arcsecond
(Wright 2006).

\begin{center}
\begin{figure*}
\includegraphics[trim=6cm 10cm 7cm 9cm, height=7cm]{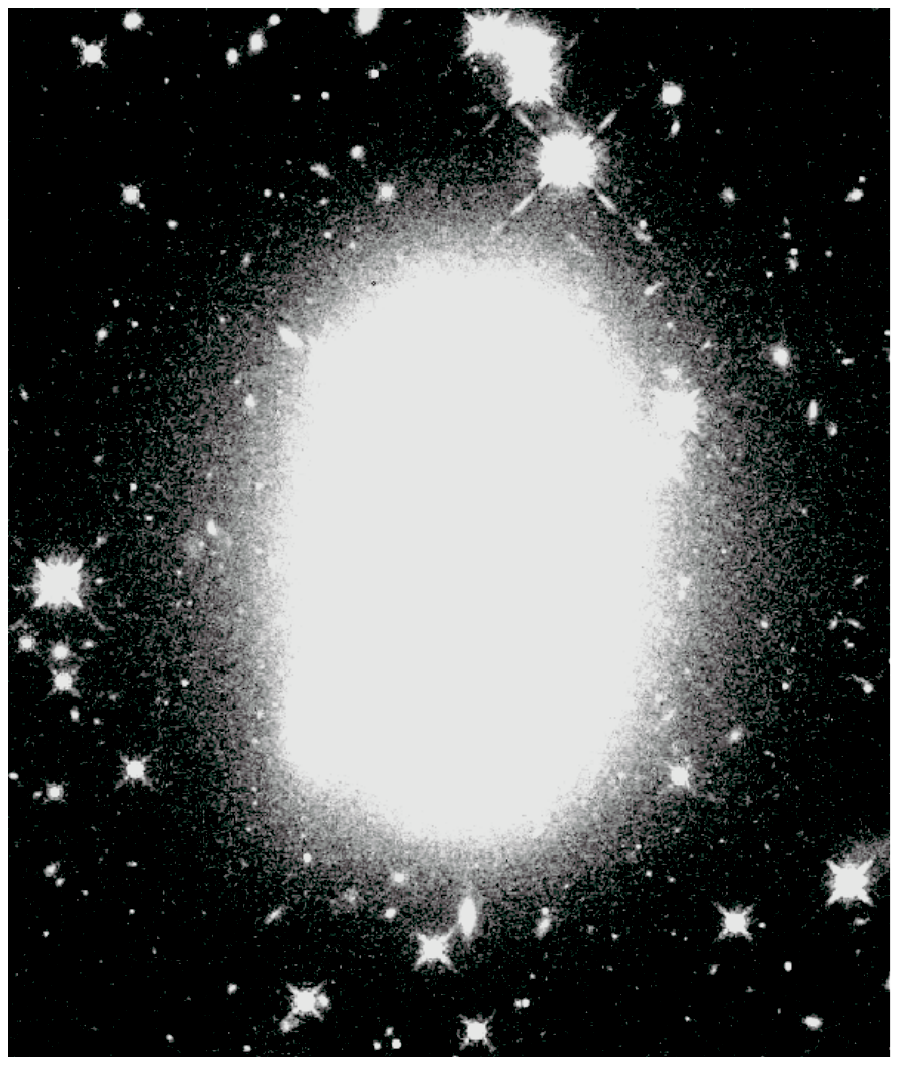}
\includegraphics[trim=-1.1cm -0.36cm 0cm +0.36cm, height=7cm]{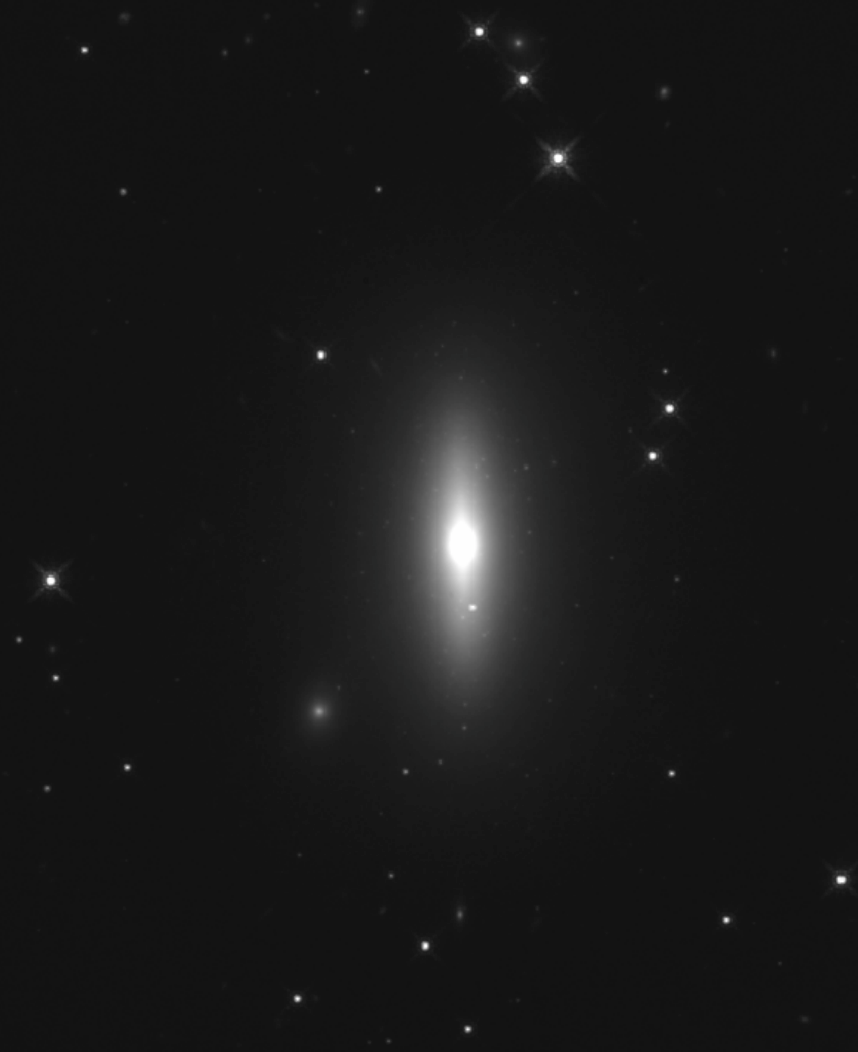}
\includegraphics[trim=6cm 10cm 6cm 9cm, height=7cm]{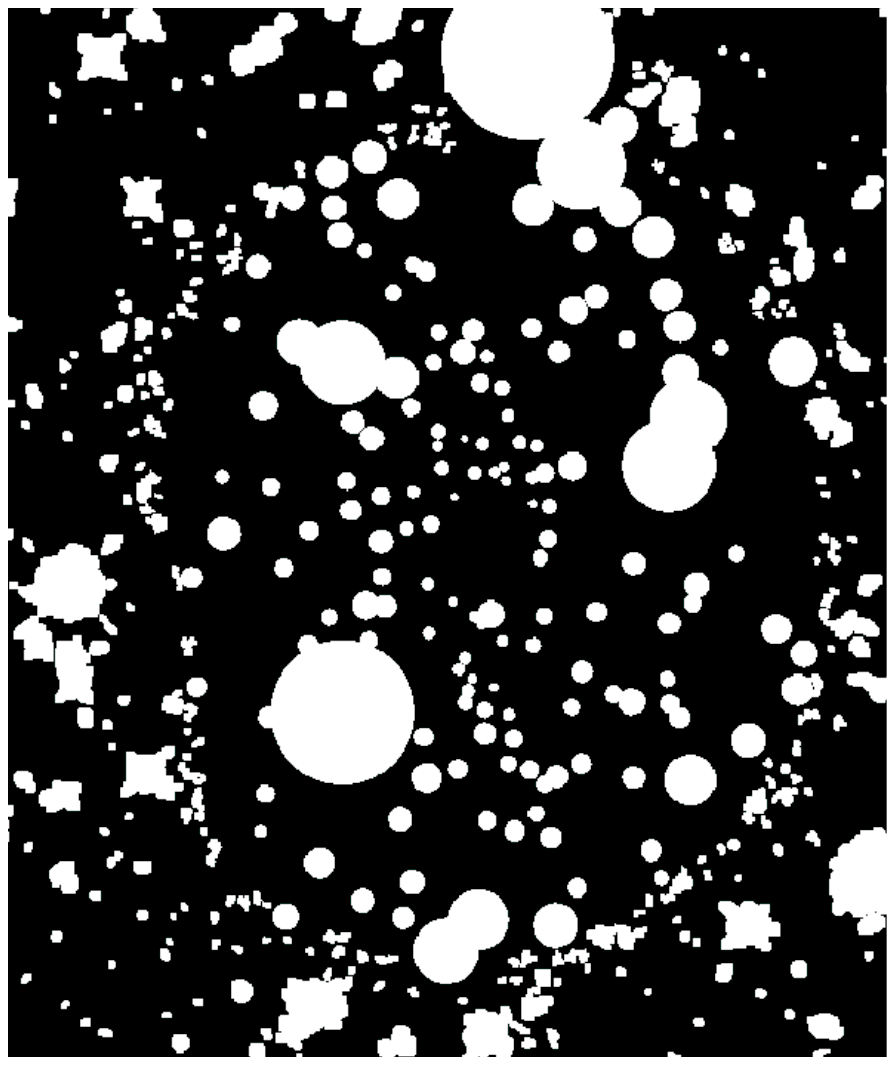}
\caption{Masking the image. 
Left panel: {\it HST/WFC3 IR/F160W} image of NGC~1271.  
Middle panel: Different stretch to show the embedded, intermediate-scale disk.
Right panel: Image mask used for subsequent galaxy modeling.
This image is $\sim$88$\arcsec$ high by $\sim$$73\arcsec$ wide. 
North is roughly toward the upper left corner while east is roughly toward the lower left corner.
}
\label{fig:image}
\end{figure*}
\end{center}

\section{Imaging Data and Analysis}\label{Sec_Intro}


We have used an archived {\it Hubble Space Telescope} ({\it HST}) image taken
with the {\it Wide Field Camera 3} ({\it WFC3}) and the near-infrared {\it F160W}
filter.  This image is a distortion corrected {\it PyDrizzle} output, obtained
from the combination of 2 individual exposures, with a total integration time
of $898.5$ seconds and a pixel size of $0.128\rm ~arcsec~pixel^{-1}$.  We
manually checked that the background subtraction had been correctly performed,
and built a mask for the contaminating sources. Figure~\ref{fig:image} shows a
cutout of NGC~1271 from the larger image, and our mask.  
NGC~1271 can be seen to contain a nearly edge-on, stellar disk. While nearly
edge-on, {\it large-scale} disks result in early-type galaxy isophotes becoming
increasingly elongated at large radii, embedded {\it intermediate-scale} disks do
not. Instead, the outer isophotes become rounder (reflecting the ellipticity of the
galaxy's spheroidal component) as the influence of the disk diminishes at
larger radii.

\subsection{1D light profile analysis}

\begin{center}
\begin{figure*}
\includegraphics[angle=0, width=\textwidth]{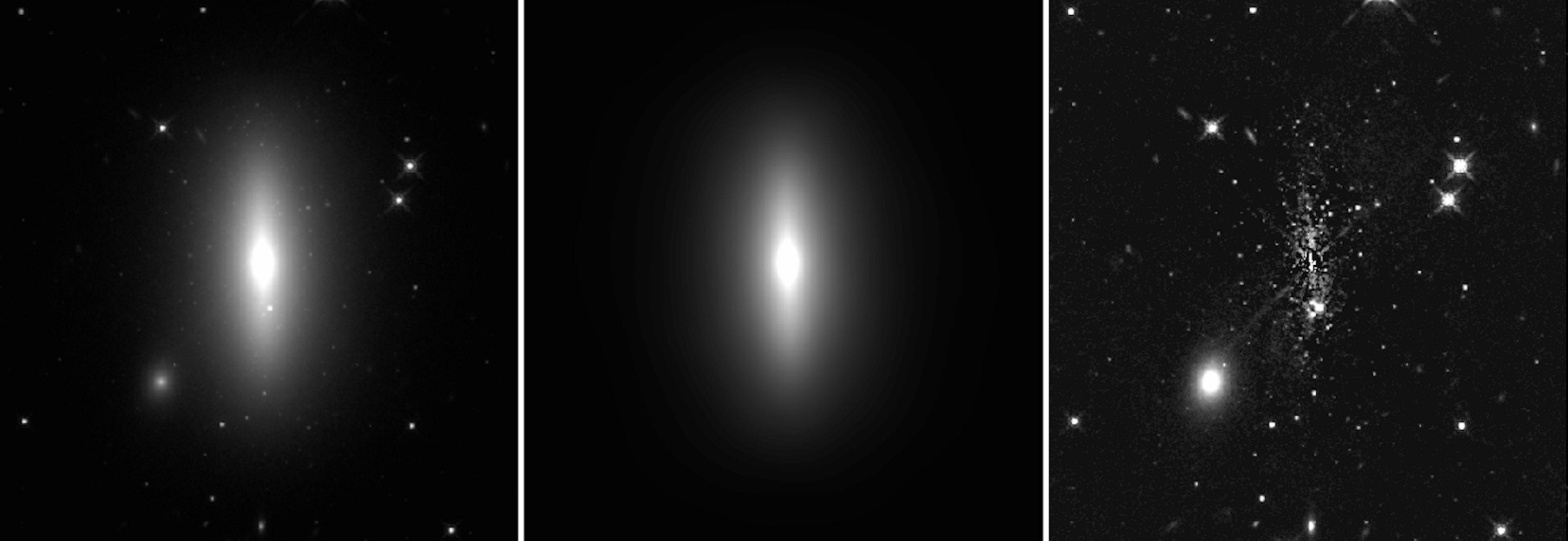}
\caption{Data, model, residual. 
Left panel: Data revealing the near edge-on disk in NGC~1271, with a 2$\times$
zoom of Figure~1. 
Middle panel: Galaxy reconstructed with the new {\sc cmodel} (construct model)
task in IRAF, 
rather than the {\sc bmodel} (build model) task, used in combination with the new 
{\sc Isofit} task rather than the {\sc Ellipse} task (see Ciambur 2015 for details). 
A similar image display stretch as in the left panel has been used. 
Right panel: Residual after subtracting the middle panel from the left panel,
and adjusting the display stretch to linear so as to effectively amplify the residuals. 
A multitude of star clusters can be seen. 
This image is $\sim$32$\arcsec$ high by $\sim$31$\arcsec$ wide.
 }
\label{fig:1D-B}
\end{figure*}
\end{center}

We performed an isophotal analysis of NGC~1271 using the task {\sc Isofit}
(Ciambur 2015), which fit quasi-elliptical isophotes to the galaxy image. {\sc
  Isofit} is a modified version of the IRAF task {\sc Ellipse} (Jedrzejewski
1987), and is much more capable of modeling features such as diskyness, which
is particularly relevant in the case of NGC~1271.  The {\sc Isofit} task generates a
major-axis light profile along with all the associated terms that describe
the shape of each isophote and enable one to construct an accurate 2D galaxy model, with 
the position angle profile, the ellipticity profile, and the Fourier harmonic terms describing
the deviations from pure elliptical isophotes at each radius. 
Our reconstructed image of NGC~1271 can be seen in the middle panel of
Figure~\ref{fig:1D-B}. As the right-hand panel of Figure~\ref{fig:1D-B}
reveals, this reconstruction provides a clean representation of the galaxy,
which is then suitable for analysis through the fitting of galaxy components.

\begin{center}
\begin{figure*}
\centering
\includegraphics[angle=0, width=0.4\textwidth]{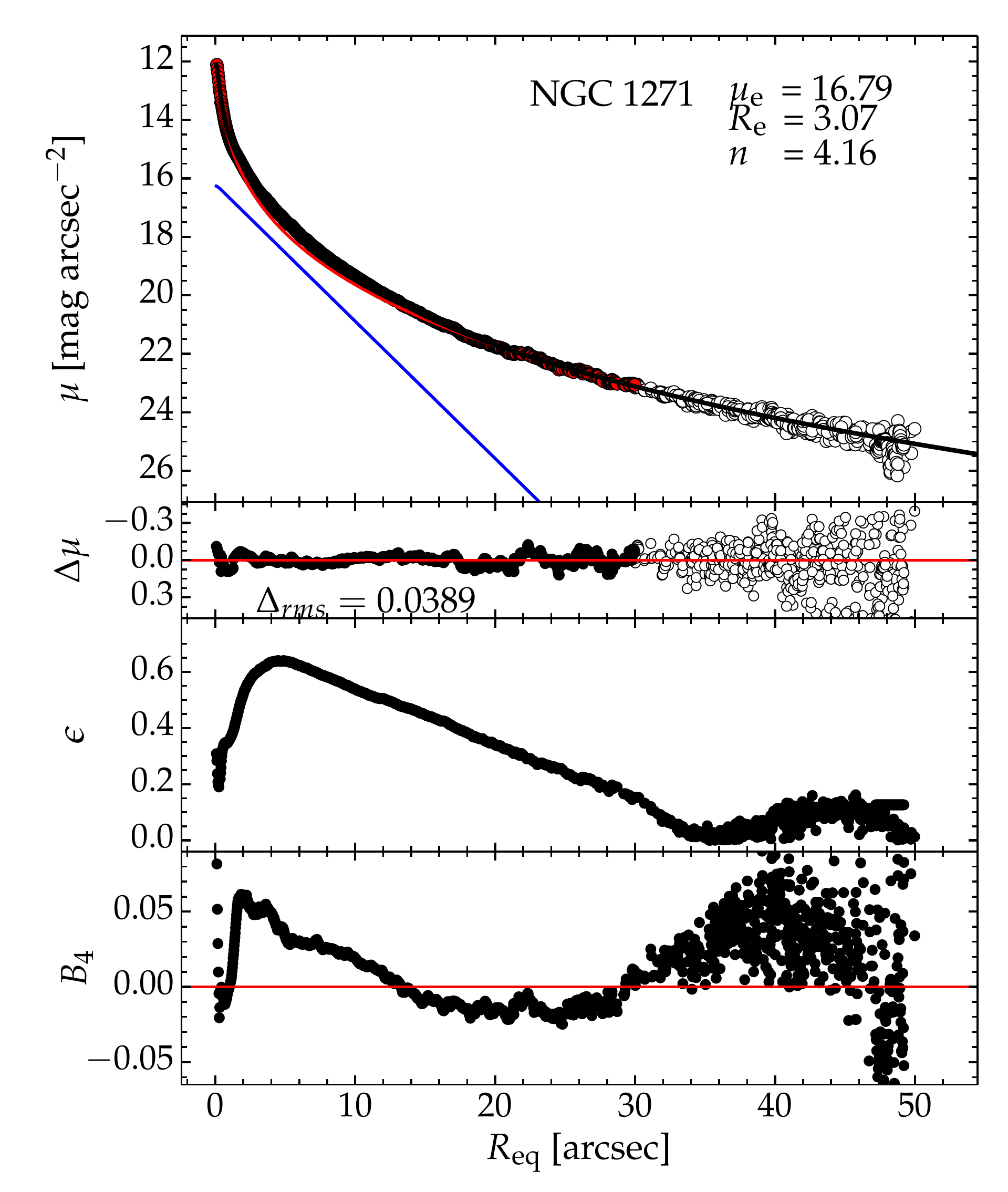}
\includegraphics[angle=0, width=0.4\textwidth]{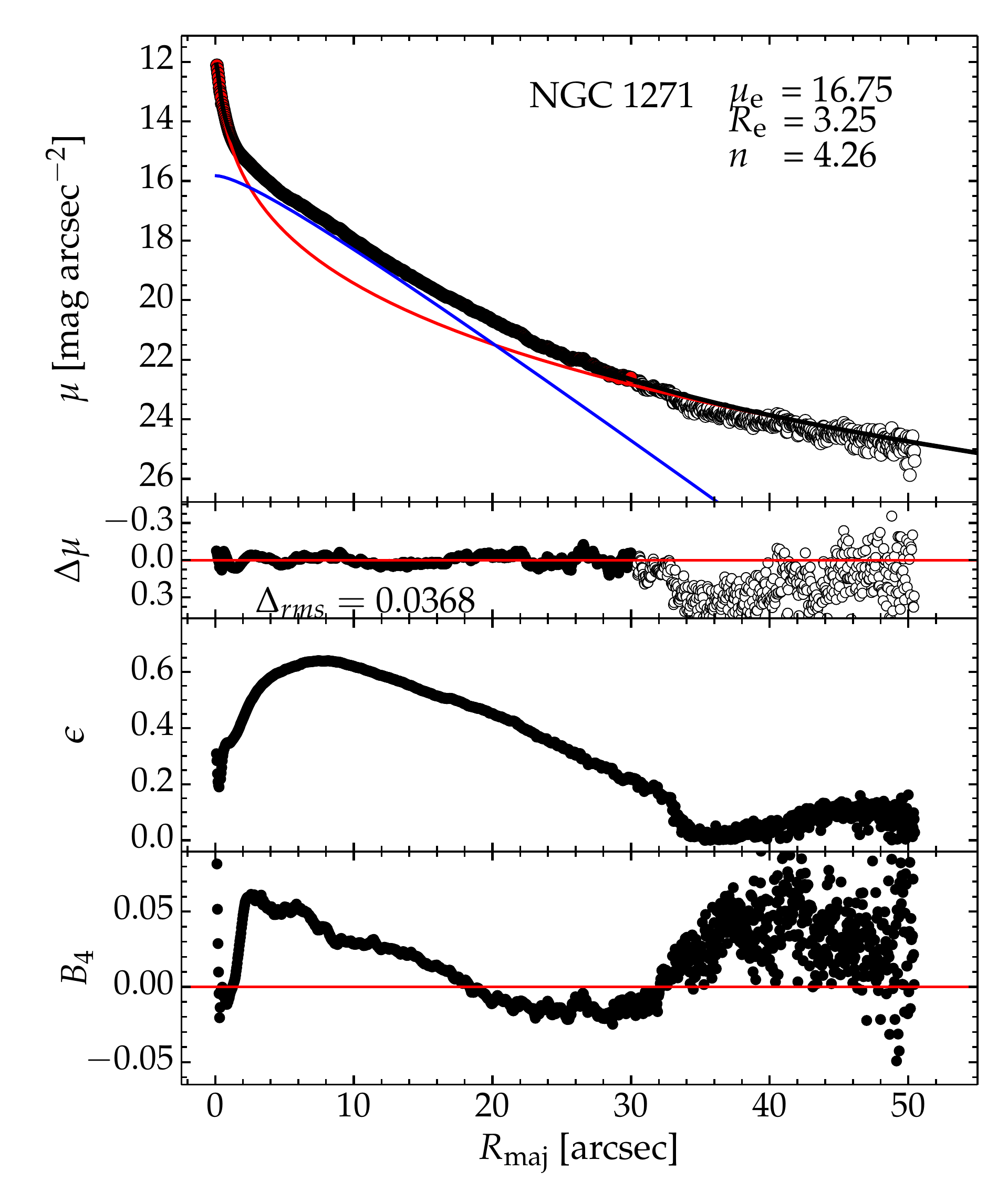}
\caption{Left panel: The geometrical mean ($\sqrt{ab}$) `equivalent axis' light
profile (calibrated to the Vega magnitude system), 
extracted from the middle panel of Figure~\ref{fig:1D-B}, is 
modeled with a S\'ersic function (red) for the galaxy's spheroidal component 
plus an exponential function (blue) for the galaxy's disk.  
Points denoted with open circles were deemed less reliable and not used in the fit. 
The S\'ersic parameters are inset in the panel, and the 
residual profile is shown in the panel immediately below.
Beneath this is the 
ellipticity and $B_4$ profiles derived using the new IRAF task {\sc Isofit}
(Ciambur 2015). 
Right panel: Major-axis ($a$) light profile, fit using
the inclined disk model (blue) and a S\'ersic function (red) for the
spheroidal component. 
}
\label{fig:1D}
\end{figure*}
\end{center}

The major-axis
surface brightness profile was additionally mapped to the so-called `equivalent-axis',
i.e.\ the geometric mean of the major ($a$) and minor ($b$) axis ($R_{\rm eq}
= \sqrt{ab}$), which is equivalent to the radius of the `circularized' isophotes.  This
axis is such that if the area interior to each isophote was enclosed by a
circle, then the radius of that circle is associated with the surface
brightness of the isophote.  With the equivalent-axis light profile one can
then use spherical geometry for the purposes of calculating the integrated surface
brightness, i.e.\ the magnitude.  Both the major-axis and the equivalent-axis
surface brightness profiles (in the Vega system) 
are shown in Figure~\ref{fig:1D}. 

The one-dimensional point spread function (PSF) was characterized using the
IRAF task {\sc Imexamine}. By fitting a Gaussian function to the radial
intensity profiles of 12 bright stars in the image, we measured the `full
width at half maximum' (FWHM) and obtained an average value of 0.24 arcsec.
We then performed a galaxy decomposition with the {\sc Profiler} software
(Ciambur 2016).  In essence, at each iteration in the fitting
process, a two-component (spheroid $+$ disk) model was constructed and
convolved with a Gaussian function to replicate the ``seeing'' (PSF) effects
in the data\footnote{Perhaps not surprisingly given the radial extent the
  light-profile, We found similar results using a Moffat function to
  represent the PSF and its Airy-ring wings.}, and then matched to the data over the inner
$30\arcsec$.  
While a core-S\'ersic profile (Graham et al.\ 2003) is expected for the
spheroidal component of NGC~1271 (see Graham \& Scott 2013) given its high
central velocity dispersion, it does not appear to have a depleted core.
NGC~1277 is another galaxy with a high central velocity dispersion and no
partially-depleted core (van den Bosch et al.\ 2012; Graham et al.\ 2016).
Indeed, rather than a deficit of stars at its core, NGC~1277 contains an
additional nuclear component.

Dullo \& Graham (2014, their section~3.4) provide a discussion of why, when
modeling the distribution of light in galaxies, one should not simply
arbitrarily add components until they are satisfied with the 
amplitude of the (data$-$model) residuals.  Several examples of where this has
led to questionable results are provided there\footnote{Readers interested in
  the structure of light profiles, and the historical development of how they
  are modeled, may wish to refer to Graham (2013).}. 
The S\'ersic (1963) $R^{1/n}$ function that we used here has previously been described in
Caon et al.\ (1993) and Graham \& Driver (2005), 
and we used a standard exponential disk model for the equivalent-axis
light profile, such that the intensity 
$I_{\rm disk}(R_{\rm eq}) = I_{0,{\rm disk}}\exp (R_{\rm eq}/h_{\rm exp})$. 
The major-axis light profile, for which the disk dominates from $\sim$3 to
$\sim$20 arcseconds, required the edge-on disk model (van der 
Kruit \& Searle 1981), which is such that the intensity 
\begin{equation}
I_{\rm disk}(R_{\rm maj}) = I_{0,{\rm disk}}\left(\frac{R_{\rm maj}}{h_{\rm edge-on}}\right)\,K_1\left(\frac{R_{\rm maj}}{h_{\rm edge-on}}\right), 
\label{Eq-edge}
\end{equation} 
where $h_{\rm edge-on}$ is a 
radial scale and $K_1$ is a modified first-order Bessel function.  
Unlike with the exponential model, $h_{\rm edge-on}$ is not the radius where
the disk intensity has dropped by a factor of e ($\approx 2.718$). 

The results of fitting the above models are shown in Figure~\ref{fig:1D},
which additionally reveals the humped ellipticity profile (and the fourth
harmonic profile: positive $B_4$ equals disky, negative $B_4$ equals boxy)
that is characteristic of the embedded disks of ES galaxies (Liller 1966).
The goodness of the fit is measured by the quantity 
$\Delta = \sqrt{ \sum_{i=1,N} ({\rm data}_i-{\rm model}_i)^2/(N-\nu)}$,
where $N$ is the number of data points and $\nu$ ($=5$) is the number of model
parameters involved (2 disk and 3 spheroid). 
The magnitudes, luminosities and ultimately stellar masses (see
Section~\ref{Subsec_Par}) of the spheroid and the embedded disk were 
derived from the equivalent-axis light profile, as discussed above and
detailed in the Appendix of Ciambur (2015). These values are provided in
Table~1.

\begin{center}
\begin{figure*}
\centering
\includegraphics[trim=0cm 0cm 0cm 0cm, width=0.33\textwidth]{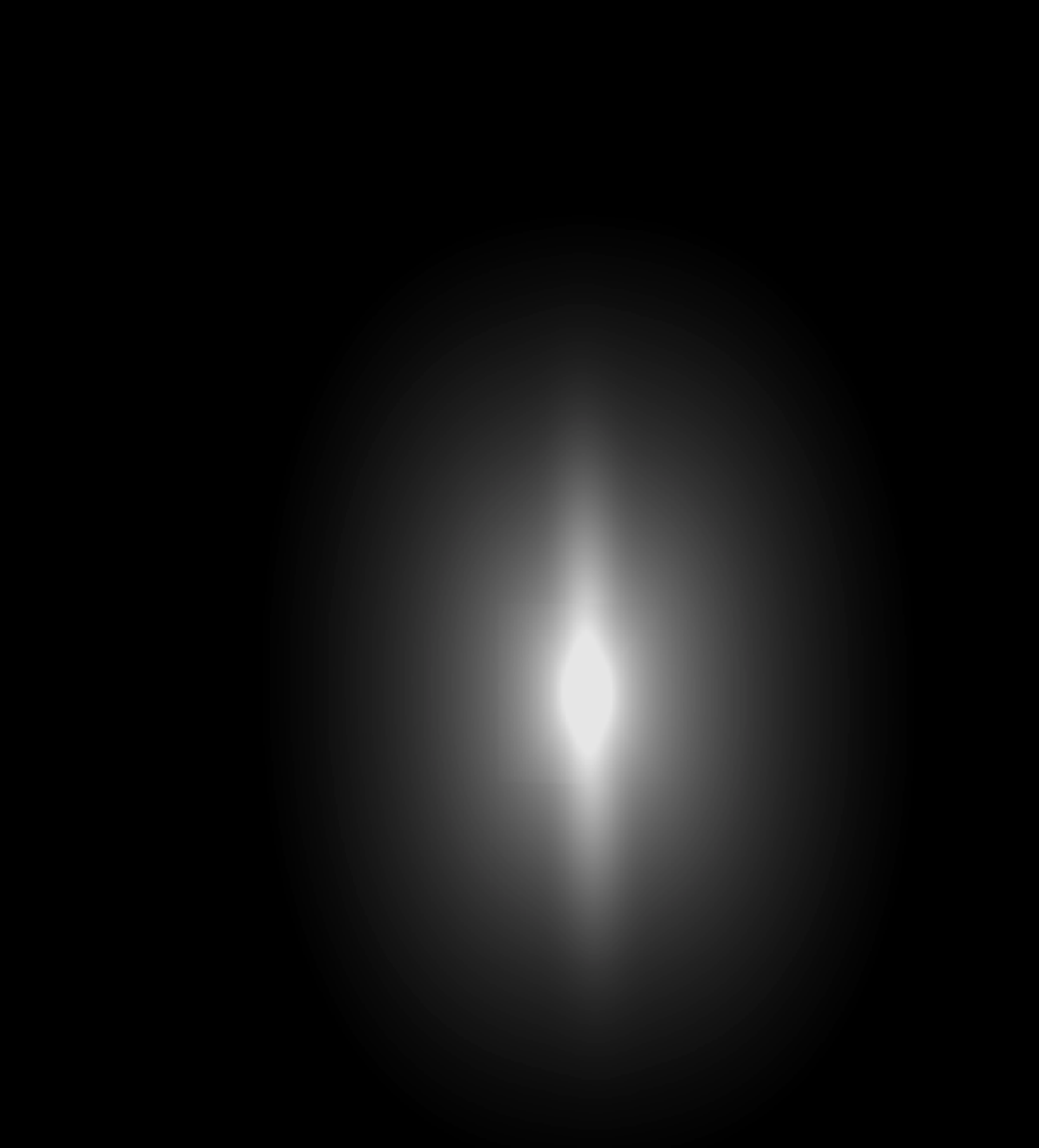}
\includegraphics[trim=0cm 0cm 0cm 0cm, width=0.33\textwidth]{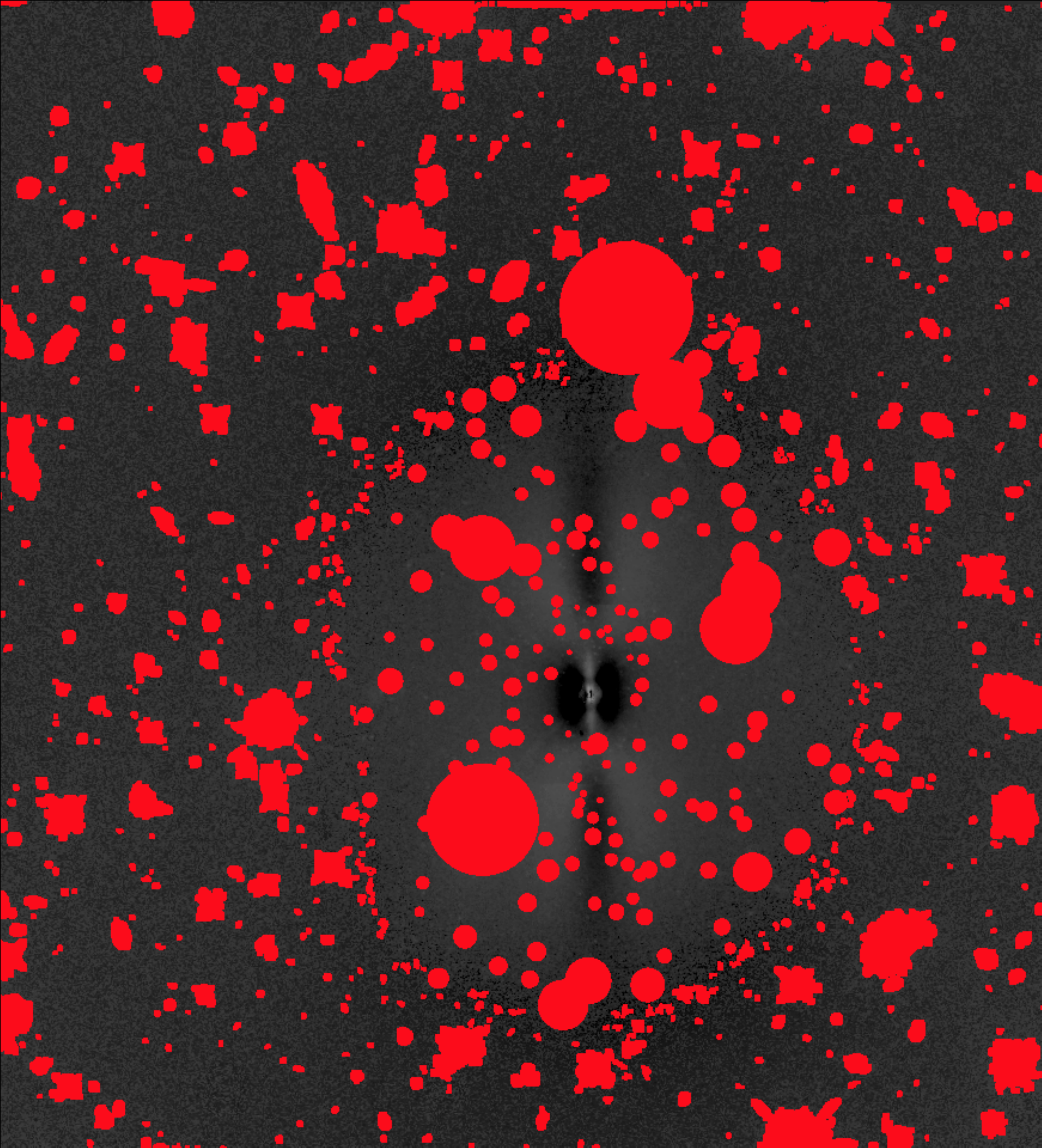}
\caption{
Left panel: 2D {\sc Galfit} model (S\'ersic bulge $+$ exponential disk) (log scaling).
Right panel: Residual image divided by the `sigma' image (greyscale, linear
scaling), and the mask used (red, linear scaling). 
This image is roughly 110$\arcsec$ high by 116$\arcsec$ wide.
 }
\label{fig:2D}
\end{figure*}
\end{center}

Given that disks have a typical thickness, i.e.\ ratio of vertical scale-height
to radial scale-length of 0.20--0.25 (e.g.\ Sandage, Freeman \& Stokes 1970;
Bizyaev et al.\ 2014, their Figure~4), we can expect the equivalent axis of a
near edge-on disk to have an exponential scale-length equal to $\sqrt{0.2}$ to
$\sqrt{0.25}$ (0.45 to 0.5) times that of the major-axis' exponential
scale-length.  The exponential disk model shown in the left hand panel of
Figure~\ref{fig:1D} has a scale-length of $2\arcsec.31$, while the edge-on disk
model shown in the right hand panel of Figure~\ref{fig:1D} has dropped in
intensity by a factor of e from its central value by the radius $R_{\rm
  maj}=5\arcsec.19$, giving a minor-to-major axis ratio of $(2.31/5.19)^2 =
0.20$.

We note that the disk is likely to be more complicated than
assumed in our models.  For instance, at around $R_{\rm
  maj}=32\arcsec$, the radial intensity profile of the disk may
truncate or break to a more rapid rate of decline (see the right hand
panel of Figure~\ref{fig:1D}). In addition, the double hump in the
$B_4$ profile at $R_{\rm maj} \approx 3\arcsec$ and $6\arcsec$ is
suggestive of another component.  There may be 
a bar or an inner ring in addition to the 
intermediate-scale disk. Furthermore, the disk component of the
equivalent-axis light profile is probably only approximately described
by the exponential model given the edge-on nature of the disk. We
trialed many variations, such as fitting equation~\ref{Eq-edge}, and
fitting the expression appropriate for the minor-axis light profile of the edge-on
disk model (van der Kruit \& Searle 1981), and locking the central
surface brightness to exactly match the value derived from the the fit
to the major-axis light profile, plus locking the scale-length to 0.45
times the asymptotic value seen at large radii in the fit to the
major-axis light profile (right hand panel of Figure~\ref{fig:1D})
where, admittedly, the disk's ``light profile'' shape is not well
constrained. Not surprisingly, none of these variations had a large
impact on the derived spheroid magnitude; the greatest departure was
just 0.15 mag. We therefore conclude that refinements to the disk
model will not have a large impact on the spheroid magnitude.

We additionally tested the robustness of our decomposition results to the PSF
characterization by using a range of FWHM between $0\arcsec .19$ --
$0\arcsec .26$ (the values reported in the literature, e.g.\ Cassata
et al.\ 2010; Windhorst et al.\ 2011).  We found this changed the 
model parameters very little: the variation in estimated bulge 
magnitude was within 0.01 mag, while the S\'ersic index along the
major-axis varied by just 0.02.

\subsection{2D image analysis}

We additionally performed a 2D image decomposition using the software {\sc Galfit3} (Peng et
al.\ 2010).  We obtained an image of the PSF from the {\sc TinyTim} software
(Krist, Hook \& Stoehr 2011; Biretta 2014).  However because this PSF is
narrower than the ``real'' stars in our image, we built a broader PSF image by
convolving a 2D Gaussian profile with the PSF image from {\sc TinyTim}.  The
advantage of this method (C.Peng, private communication) is to obtain a 2D PSF
that is wider than the {\sc TinyTim}
PSF, but maintains the asymmetric features of the {\sc TinyTim} PSF (e.g.\ wings and
spikes).   We modeled the galaxy with a S\'ersic component for the spheroid
and an exponential function for the disk. The minor-to-major axis ratio for
the fitted disk model was equal to 0.22.  As noted above, this is basically equal to the
typical thickness of galaxy discs, i.e.\ the 
(vertical scale-height)-to-(radial scale-length) ratio observed in edge-on
discs.  We note that using a S\'ersic model
for the edge-on disk, rather than an exponential model, had no significant
impact on the results (Figure~\ref{fig:2D} and Table~1). 

Our 2D analysis is similar to the 2-component fit
reported in Table~4 of W2015, in that the disk component has similar
parameters.  While we also both report a S\'ersic index $\sim$6 for the more
massive component, W2015 obtain notably more elongation from their fit, such
that they report an axis ratio of 0.54 for their S\'ersic model, compared to our
value of 0.69, and they report a major-axis half-light radius of $R_{\rm e} =
5\arcsec .23$ for the  spheroid compared to our value of 3$\arcsec$.08.  An axis ratio
of 0.54 is too low to match the outermost ellipse drawn in Figure~1 of W2015,
which approximates the isophote at $R_{\rm major-axis}=25\arcsec$ and has an
axis ratio of 0.65.  

The rigidity of the 2D model's components (i.e.\ the constant
position angle and ellipticity of the spheroid and disk), 
and the treatment of the diskyness, 
meant that our 2D model was not able to fully reproduce the 
galaxy light in the way that the 1D analysis did. 
This is evident in the residual image seen in the right hand panel of
Figure~\ref{fig:2D}, which can be contrasted with the right hand panel of 
Figure~\ref{fig:1D-B}. 
Despite the better result using the 1D approach, the difference in
the magnitude for both the spheroid and the disk is not more than 0.3 mag when using 
the 2D approach.  More specifically, the 2D model results in a spheroid that
is just 22\% fainter (see Table~1).  Given that the 1D light 
profile is constructed by fully capturing the galaxy light (see
Figure~\ref{fig:1D-B}), while also accounting for deviations from elliptical
isophotes in terms of Fourier harmonics, this is our preferred
analysis for the estimate of the spheroid brightness. 
Fitting the 1D light profile, within the inner 30$\arcsec$, the difference between our
2-component spheroid$+$disk model and the data is less than 0.15 mag
arcsec$^{-2}$, while the average difference is less than 0.04 mag 
arcsec$^{-2}$ (Figure~\ref{fig:1D}).

\begin{table*}[ht]
\centering
\label{Tab_Comp}
\caption{Galaxy model's component parameters.}
\begin{tabular}{lccccccccc} 
\hline\hline
Component & $R_{\rm e}$      & $h_e$          & $n$   & $(b/a)$  & $m_{\rm H}$ & $M_{\rm H}$  & $L_{\rm H}$ & $M_*/L_H$           & $M_{\rm *}$     \\ 
          & (arcsec / kpc) & (arcsec / kpc) &       &          & (mag)     & (mag)      & $10^{10}\rm ~L_{\odot}$ & $M_{\odot}/L_{\odot}$ & $10^{10}\rm ~M_{\odot}$ \\ 
 (1)      &   (2)          &  (3)           & (4)   &  (5)     &  (6)      &  (7)       &   (8)      &   (9)   & (10)   \\ 
\hline                                                                                                                         
\multicolumn{9}{c}{\emph{2D modeling of the image} (major-axis scale sizes)} \\  
Spheroid  &  3.08 / 1.17   &  ...           &  5.84 &  0.69    &  11.22    &  $-23.42$  &   5.01    &  1.4     &  7.01   \\ 
Disk      &  ...           &  3.99 / 1.51   &  1.00 &  0.22    &  12.08    &  $-22.56$  &   2.27    &  1.4     &  3.18   \\ 
\multicolumn{9}{c}{\emph{1D modeling of the major-axis light profile}} \\
Spheroid  &  3.25 / 1.23   &  ...           &  4.26 &  ...     &  ...      & ...        &    ...    &  ...     &  ...    \\ 
Disk      &  ...           &  5.19 / 1.97   &  1.00 &  ...     &  ...      & ...        &    ...    &  ...     &  ...    \\ 
\multicolumn{9}{c}{\emph{1D modeling of the equivalent-axis light profile}} \\
Spheroid  &  3.07 / 1.16   &  ...           &  4.16 &  1.00    &  10.95    &  $-23.69$  &   6.43    &  1.4     &  {\bf 9.00} \\ 
Disk      &  ...           &  2.31 / 0.88   &  1.00 &  1.00    &  12.38    &  $-22.26$  &   1.72    &  1.4     &  2.41   \\ 
\hline\hline
\end{tabular}
\\
Column~1: Model component that was fit to the light distribution.
Column~2: Effective half light radius. 
Column~3: Radius where the disk's intensity is
e ($\approx 2.718$) times fainter than its central intensity (Note: the edge-on disk model
used to describe the major-axis light profile has a radial scale $h_{\rm edge-on}$ equal to $3\arcsec.13$.)
Column~4: S\'ersic index. 
Column~5: Component axis-ratio. 
Column~6: Observed (uncorrected) $H$-band apparent magnitude (Vega). 
Column~7: (Corrected, by 0.12 mag, see the text for details) $H$-band absolute magnitude. 
Column~8: $H$-band solar luminosity.
Column~9: $H$-band Stellar mass-to-light ratio. 
Column~10: Stellar mass.
\end{table*}

\begin{center}
\begin{figure*}
\centering
\includegraphics[trim=3cm 2.3cm 3.3cm 9cm, angle=0, width=0.7\textwidth]{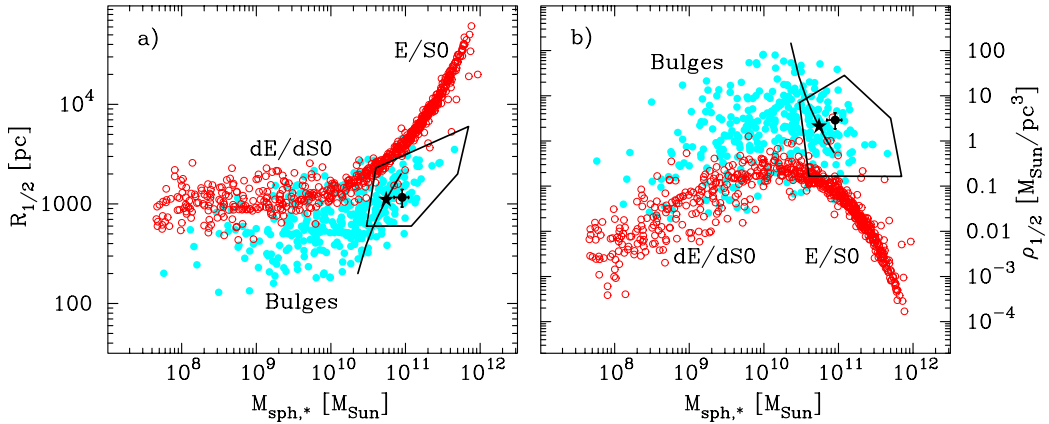}
\caption{Panel a) Projected half-light radius ($R_{1/2}\equiv R_{\rm e}$) 
versus stellar mass for nearby early-type galaxies 
and the bulges of nearby spiral, lenticular and ellicular galaxies. 
Panel b) Internal stellar-mass density within the
radius ($4/3 \times R_{1/2}$) containing half the stellar mass plotted against
the total stellar mass.  
Our estimate for NGC~1271 is shown by the black dot, while 
the range in mass and radii according to W2015 is shown by the black curve,
with their midpoint marked by the black star. 
The distribution of compact massive galaxies at $z\sim1.5$ --- as given by
Damjanov et al.\ (2009, their Figure~5) --- is denoted by the 5-sided boundary
which can be seen to overlap with the massive bulges at $z\approx0$ shown
here.  Adapted from Figure~1 in Graham (2013). 
 }
\label{fig:ab}
\end{figure*}
\end{center}

\newpage

\subsection{Galaxy parameters}\label{Subsec_Par}

Table~1 lists the structural parameters that we have obtained for NGC~1271. 
We correct the observed, apparent magnitudes of each component for 2.5$\log(1+z)^2$
($\approx$0.04) mag of cosmological redshift dimming, and 0.08 mag of Galactic
extinction (Schlafly \& Finkbeiner 2011, via the NASA/IPAC Extragalactic
Database: NED\footnote{http://nedwww.ipac.caltech.edu}).
The intermediate-scale disk of NGC~1271 does not appear to contain any obvious
dust features beyond the inner 1$\arcsec$, and 
as such we have elected to not apply an internal dust correction which would
act to brighten the disk and spheroid (Driver er al.\ 2008). 

Using a distance modulus\footnote{Based on a distance of 80 Mpc.} of
34.52, and a solar $H$-band absolute magnitude of $+$3.33 mag (Vega:
Bessell et al.\ 1998), the corrected absolute magnitude of the
spheroid ($-23.71$ mag) corresponds to an $H$-band luminosity of
$6.43\times 10^{10}~L_{\odot}$.  W2015 report an $H$-band stellar
mass-to-light ratio of $1.40^{+0.13}_{-0.11}$ for NGC~1271, and thus
we have a stellar mass of $9.00\times 10^{10}~M_{\odot}$ for the
spheroidal component of NGC~1271. Coupling the 0.15 magnitude
uncertainty on the spheroid magnitude derived from the 1D light profile
analysis with the above uncertainty on the
mass-to-light ratio, we roughly have that 
$M_{\rm sph,*} = (9\pm2)\times 10^{10}\rm~M_{\odot}$. 

Our spheroid-to-disk mass ratio is 3.7 (see Table~1). 
For reference, the 2-component fit by W2015 reported a spheroid-to-disk mass 
ratio equal to 4.1 and a bulge mass of $8.7\times 
10^{10}~M_{\odot}$, in excellent agreement with our results. 
While that mass was their upper limit, the value of $9\times 10^{10}\rm~M_{\odot}$
is our nominal value.  
W2015 also fit a 3-component model to NGC~1271,
with the third component having a major-axis half-light radius of
0$\arcsec$.61 and a stellar mass of $2.2\times 10^{10}~M_{\odot}$. They
thought that this might be the actual ``bulge'' (i.e.\ the spheroid) to be
used in the $M_{\rm bh}$--$M_{\rm sph}$ scaling relation.  As they were unsure
as to the spheroidal component of this massive early-type galaxy, they
averaged these two masses and proceeded by assuming a mean spheroid mass of
$5.4\times10^{10}~M_{\odot}$. 

The location of NGC~1271's spheroid, as determined by our analysis, 
is shown in the mass-size and mass-density 
diagrams in Figure~\ref{fig:ab}.  It appears to be a common object,
over-lapping with the physical properties of other massive bulges.  An
expression for the curved relation describing the distribution of the 
early-type galaxies in Figure~\ref{fig:ab}a can be found in Graham (2013, 
his section~3.2.2; see also Lange et al.\ 2015, their Figure~6).

\subsection{Predicted Black hole mass}\label{subSec_Pred}

We predict the black hole mass in NGC~1271 using three relations related
to the spheroidal component of galaxies (Table~2).

Using the stellar mass of the spheroidal component of NGC~1271, 
the `S\'ersic' $M_{\rm bh}$--$M_{\rm sph,*}$ relation in Scott et al.\ 
(2013), is such that 
\begin{eqnarray}
\log(M_{\rm bh}/M_{\odot}) & = & (7.89\pm0.18)  \nonumber \\ 
 & & \hskip-40pt + (2.22\pm0.58)\log 
       \left[ M_{\rm sph,*}/2\times 10^{10}~M_{\odot} \right]. 
\label{Eq_MM}
\end{eqnarray}
Given $M_{\rm sph,*} = 9.00 \times 10^{10}~M_{\odot}$, and assigning a
20\% uncertainty to this value\footnote{See section~3.3 in Graham \& Scott
  (2013) to understand how this uncertainty propagates through to the
  predicted black hole mass. Following that study, an intrinsic scatter of
  0.90 dex in the $\log(M_{\rm bh})$ direction has been used here for
  equation~\ref{Eq_MM}.}, 
one has the prediction 
$\log(M_{\rm bh})=9.34 \pm 1.01$ 
($M_{\rm bh} = 2.19\times 10^9~M_{\odot}$ with an order or magnitude uncertainty). 
This agrees well with the reported measurement from W2015 
of $(3.0^{+1.0}_{-1.1})\times 10^9~M_{\odot}$. 
The large uncertainty in the estimated value is dominated by the 0.90 dex of
intrinsic scatter in the $\log(M_{\rm bh})$ direction about that 
S\'ersic $M_{\rm bh}$--$M_{\rm sph,*}$ relation (Scott et al.\ 2013). 

For a black hole of this mass, residing in a spheroid with a velocity
dispersion $\sigma = 276$ km s$^{-1}$ (W2015), its sphere of influence $r_h
\equiv GM_{\rm bh}/\sigma^2 = 0\arcsec.32$.  For reference, the Near-infrared
Integral Field Spectrometer (NIFS; McGregor et al.\ 2003) data used by W2015
to determine the black hole mass had a point-spread function which was
represented by the sum of two Gaussians with dispersions of $0\arcsec.16$ and
$0\arcsec.43$ and a 0.61:0.39 Strehl ratio weighting.

Since submitting this work, a new relationship for early-type galaxies has
become available (Savorgnan et al.\ 2016), and is such that 
\begin{eqnarray}
\log(M_{\rm bh}/M_{\odot}) & = & (8.56\pm0.07)  \nonumber \\ 
 & & \hskip-40pt + (1.04\pm0.10)\log 
       \left[ M_{\rm sph,*}/10^{10.81}~M_{\odot} \right], 
\label{Eq_Set}
\end{eqnarray}
with an intrinsic scatter ($\sigma_{\rm intrinsic}$) 
of about 0.48 dex in the  $\log(M_{\rm bh})$ direction. 
This relationship was established using the careful galaxy decompositions
presented in Savorgnan \& Graham (2016a), which revealed problems with many past
works. 
Using $M_{\rm sph,*} = 9 \times 10^{10}~M_{\odot}$, 
this relation gives $\log(M_{\rm bh})=8.71 \pm 0.49$, or 
$M_{\rm bh} = (0.51^{+1.07}_{-0.35})\times 10^9~M_{\odot}$ (1-sigma uncertainties). 
While this is 5.9 times lower than the reported black hole mass
of $3\times 10^9~M_{\odot}$, the
upper 2-sigma bound on this estimated mass is $4.9\times
10^9~M_{\odot}$, encompassing the reported black hole mass.
The reported black hole mass is offset by 0.77 dex from equation~\ref{Eq_Set}, or 1.6
times $\sigma_{\rm intrinsic}$.  Refining the scatter about the $M_{\rm
  bh}$--$M_{sph,*}$ relation will of course refine the significance of this 
offset. 

From a sample of 89 galaxies, Savorgnan \& Graham (2015) present an updated 
(black hole mass)--(velocity dispersion: $\sigma$) diagram. 
It accounts for offset barred galaxies,
i.e.\ the substructure in the $M_{\rm bh}$--$\sigma$ diagram (Graham
2008; Hu 2008) and reveals that the $M_{\rm bh}$--$\sigma$ relation has not
saturated at high $\sigma$ values due to increased dry mergers 
(at odds with Volonteri \& Ciotti 2013, and 
Kormendy \& Ho 2013, their section~6.7).  Excluding 
the barred galaxies, equation~1 from Savorgnan \& Graham (2015) is such that 
\begin{eqnarray} 
\log(M_{\rm bh}/M_{\odot}) &  = & (8.24\pm0.10) \nonumber \\
 & & \hskip-40pt + (6.34\pm0.80)\log \left[ \sigma/200 {\rm ~km~s}^{-1} \right], 
\label{Eq_Ms}
\end{eqnarray}
where $\sigma$ is typically the value within the inner few arcseconds.
W2015 report a `bulge' stellar velocity dispersion of 276 km s$^{-1}$
within $2.\arcsec 9$ (when excluding the inner $0.\arcsec 44$
``sphere-of-influence'' based on their directly measured black hole mass
of $3\times 10^9~M_{\odot}$).  This velocity dispersion gives a
predicted black hole mass\footnote{The uncertainty on this black hole
  mass was derived assuming 0.3 dex of intrinsic scatter in the
  $\log(M_{\rm bh})$ direction of the $M_{\rm bh}$--$\sigma$ diagram
  and a 10\% uncertainty on NGC~1271's adopted velocity dispersion.}
of $\log(M_{\rm bh})=9.13 \pm 0.43$, or $M_{\rm bh} =
(1.35^{+2.28}_{-0.85}) \times 10^9~M_{\odot}$.  Therefore, the reported
black hole mass in NGC~1271 is in agreement with expectations from the
$M_{\rm bh}$--$\sigma$ relation, as reported by W2015.

\begin{table}
\label{Tab_BH}
 \centering
\caption{Predicted black hole mass.}
\begin{tabular}{lcc}
\hline\hline
Relation & Parameter                               & $M_{\rm bh}$ prediction  \\
         &                                               & ($10^9~M_{\odot}$)  \\
\hline
S\'ersic $M_{\rm bh}$--$M_{\rm sph,*}$    & $(9\pm2)\times 10^{10}~M_{\odot}$  & $2.19^{+20.19}_{-1.98}$  \\
Early-types $M_{\rm bh}$--$M_{\rm sph,*}$ & $(9\pm2)\times 10^{10}~M_{\odot}$  & $0.51^{+1.07}_{-0.35}$  \\    
Non-barred $M_{\rm bh}$--$\sigma$         &    276 km s$^{-1}$                 & $1.35^{+2.28}_{-0.85}$ \\
\hline
\end{tabular} 

The relations are presented in Section~\ref{subSec_Pred}.
\end{table}

\section{Discussion}\label{Sec_Disc}

\subsection{$M_{\rm bh}$--$M_{\rm sph}$  relations}

As revealed in Graham (2012) and Graham \& Scott (2013), the $M_{\rm 
  bh}$--$M_{\rm sph}$ distribution is bent; it is not described by a single
log-linear relation. 
A number of simulations of galaxy evolution display a bend in the
$M_{\rm bh}$--$M_{\rm sph}$ diagram, with some performing 
better than others at quantitatively matching the observed break and slopes 
(e.g.\ Cirasuolo et al.\ 2005, their Figure~5; 
Fontanot et al.\ 2006, their Figure~6; 
Dubois et al.\ 2012, their Figure~3; 
Khandai et al.\ 2012, their Figure~7; 
Bonoli, Mayer \& Callegari 2014, their figure~7; 
Lu \& Mo 2015;  
Fontanot et al.\ 2015). 
While the slope of the ``core-S\'ersic $M_{\rm bh}$--$M_{\rm sph}$ relation'' is
roughly linear, the $M_{\rm bh}$--$M_{\rm sph}$ relation is steeper for the
S\'ersic spheroids, i.e.\ those without partially depleted cores (Graham \&
Scott 2013) --- perhaps
largely due to the inclusion of 
the bulges of spiral galaxies (Savorgnan et al.\ 2016). 

Returning to NGC~1271, with its intermediate-scale disk, W2015 claimed
that its directly measured black hole mass of
$(3.0^{+1.0}_{-1.1})\times 10^9~M_{\odot}$ is an order of magnitude
larger than their expectation from the near-linear $M_{\rm bh}$--$M_{\rm
  sph,*}$ relation that they used from Kormendy \& Ho (2013).  After
identifying the spheroidal component of NGC~1271 and
obtaining a stellar mass of $9\pm2 \times 10^{10}~M_{\odot}$, we find
that the measured black hole mass is a factor of 6.9 above the
predicted black hole mass of $4.34\times10^8~M_{\odot}$ when using the
relation from Kormendy \& Ho (2013, their equation~10).  Savorgnan \&
Graham (2016a) have since provided superior image decompositions for galaxies
with directly measured black hole masses, and
a clear explanation of problems with many past decompositions.  From
their improved near-infrared magnitudes for the spheroidal components 
of these galaxies, Savorgnan et
al.\ (2016) present an updated $M_{\rm bh}$--$M_{\rm sph,*}$ relation
for the early-type galaxies.  As noted in Section~\ref{subSec_Pred}, this yields a consistent black hole mass of
$(5.1^{+10.7}_{-3.5})\times10^8~M_{\odot}$ (1-sigma uncertainties).
However, the 2-sigma uncertainty ranges from
$0.54\times10^8~M_{\odot}$ to $4.9\times10^9~M_{\odot}$, encompassing
the reported black hole mass of $3\times10^9~M_{\odot}$.  The reported mass is
just 1.6 standard deviations 
away from the $M_{\rm bh}$--$M_{\rm sph,*}$ relation for early-type
galaxies\footnote{The offset is even less when using the total rms scatter
  rather than the intrinsic scatter associated with equation~\ref{Eq_Set}.}.

It is of interest to compare the $M_{\rm bh}/M_{\rm sph,*}$ mass ratio
in NGC~1271 with other massive galaxies.  Graham (2012) reported that
massive spheroids with partially depleted cores (known as
core-S\'ersic galaxies) have an average $M_{\rm bh}/M_{\rm sph,dyn}$
mass ratio of 0.36\% and, using a larger data sample, Graham \& Scott
(2013) reported an average $M_{\rm bh}/M_{\rm sph,*}$ mass ratio of
0.49\%.  Kormendy \& Ho (2013) subsequently reported the same ratio at
$M_{\rm sph,*}=10^{11}~M_{\odot}$, and the dashed-line in
Figure~\ref{fig:M-M} (showing their scaling relation) largely reproduces the core-S\'ersic relation
from Scott et al.\ (2013) at high masses.  The $M_{\rm bh}/M_{\rm sph,*}$ mass ratio
in NGC~1271 is $(3.0\times 10^9)/(9.0\times 10^{10})$, equal to 3.3\%.

\begin{center}
\hspace{-1cm}
\begin{figure}
\centering
\includegraphics[trim=2cm 0cm 2cm 0cm, angle=0, width=0.85\columnwidth]{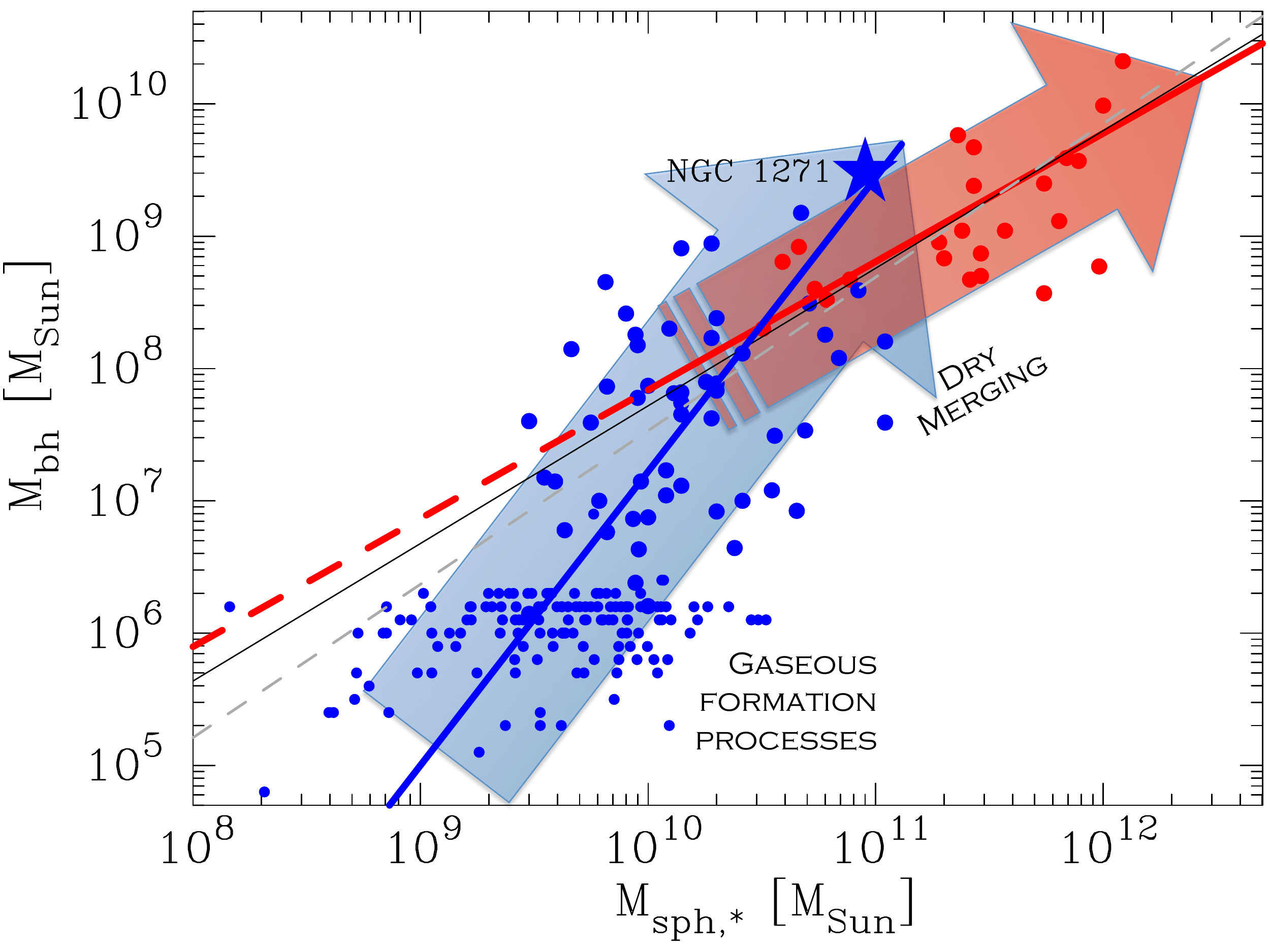}
\caption{Mass tracks.  The $M_{\rm bh}$--$M_{\rm sph,*}$ diagram from Graham
  \& Scott (2015), with the addition here of NGC~1271 (denoted by the star).
  Small blue dots denote AGN with $M_{\rm bh} < 2\times10^6~M_{\odot}$, while
  larger blue dots denote S\'ersic spheroids, and red dots represent
  core-S\'ersic spheroids.  The near-linear (log-slope $=0.97\pm0.14$) and
  near-quadratic (log-slope $=2.22\pm0.58$) scaling relations from Scott et
  al.\ (2013) are shown as the thick red (solid and dashed) and thick blue
  (solid) line
  for the core-S\'ersic and S\'ersic spheroids, respectively.  The AGN and
  NGC~1271 were not included in the derivation of these scaling relations.
  The thin black line is the relation for early-type galaxies from Savorgnan
  et al.\ (2016), and 
  the thin dashed grey line is the relation from Kormendy \& Ho (2013). 
  }
\label{fig:M-M}
\end{figure}
\end{center}

Rather than building a large-scale disk, perhaps ES galaxies with
their intermediate-scale disks used their gas supplies to keep building
their spheroid and black hole such that they continued to evolve along
the steeper $M_{\rm bh}$--$M_{\rm sph,*}$ scaling relation for
S\'ersic galaxies (see Figure~\ref{fig:M-M}).  
Such a scenario was described in Section~4.4 of 
Graham \& Scott (2015).  However, the rms scatter about the
``core-S\'ersic $M_{\rm bh}$--$M_{\rm sph,*}$ relation'' from Graham \&
Scott (2013) is 0.47 dex in the $\log(M_{\rm bh})$ direction, in close
agreement with the scatter about the $M_{\rm bh}$--$M_{\rm sph,*}$
relation for early-type galaxies from Savorgnan et al.\ (2016), which
contained less S\'ersic galaxies than Graham \& Scott (2013).  We {\it can}
therefore expect to find some galaxies with mass ratios that are
$\pm2\sigma_{\rm intrinsic}$ ($\pm0.94$ dex, i.e.\ a factor of 8.7) higher and lower
than the core-S\'ersic $M_{\rm bh}$--$M_{\rm sph,*}$ relation.  The
reported black hole mass in NGC~1271 is thus consistent with the
near-linear $M_{\rm bh}$--$M_{\rm sph,*}$ relations.
That is, it would be premature to claim NGC~1271 as an example of an
over-massive black hole in a S\'ersic galaxy following the steeper $M_{\rm
  bh}$--$M_{\rm sph,*}$ relation and deviant from the near-linear relation. 
The high-$z$ galaxy reported by Trakhtenbrot et al.\ (2015) may however evolve into
such an example. 

While the location of NGC~1271 in the $M_{\rm bh}$--$M_{\rm sph,*}$ diagram
resides almost exactly on the steep S\'ersic relation from Scott et al.\ (2013), it should be
kept in mind that this relation does contain scatter, and one data point
(NGC~1271) is prone to chance alignment. 
If other 
massive early-type galaxies with intermediate-scale disks (thus keeping the galaxy compact like
its spheroid) and without partially depleted cores, are found to have high
$M_{\rm bh}$/$M_{\rm sph,*}$ ratios like NGC~1271, then collectively they are unlikely to
all be {\it positive} outliers from the near-linear core-S\'ersic 
relation.  Instead, they may reveal how the S\'ersic $M_{\rm bh}$/$M_{\rm sph,*}$
relation continues to higher masses.  Preliminary 
reports of {\it elevated} $M_{\rm bh}$/$M_{\rm sph,*}$ ratios (relative 
to expectations from the near linear $M_{\rm bh}$--$M_{\rm sph,*}$ relation)
in other massive S\'ersic galaxies by Ferr\'e-Mateu et al.\ (2015) may
therefore not actually 
challenge the co-evolution of super-massive black holes.  Rather, 
their host spheroid may simply be compliant with expectations from the 
S\'ersic $M_{\rm bh}$--$M_{\rm sph,*}$ relation defined by less massive
S\'ersic spheroids.  However, as done here with NGC~1271, and in Graham et
al.\ (2016) with NGC~1277, the spheroid masses first need to be reliably
derived.
This has now been done for NGC~1332, NGC~3115 and Mrk~1216 
 (Savorgnan \& Graham 2016b)\footnote{At the time of submission (9
  July, 2015), Savorgnan \& Graham (2016b) was work in prep.}
and NGC~821, NGC~3377 and NGC~4697 (Savorgnan \& Graham 2016a).

\subsection{Compact massive spheroids - evolution or not}

For decades the presence and influence of (somewhat face-on) 
large stellar disks in nearby early-type galaxies 
has frequently been over-looked.  Capaccioli (1987, 1990), Carter (1987),
and Capaccioli et al.\ (1990) campaigned for a greater awareness of these disks
(see also Poulain et al.\ 1992 in the case of the Perseus cluster of galaxies
containing NGC~1271 and NGC~1277), and the growing volume of kinematical data
has since confirmed the prevalence of such disks (e.g.\ Nieto et al.\ 1988;
D'Onofrio et al.\ 1995; Graham et al.\ 1998; Emsellem et al.\ 2011; Scott et
al.\ 2014). Indeed, Emsellem et al.\ (2011, see their Figure~11), through the ATLAS$^{\rm 3D}$
project (Cappellari et al.\ 2011a), revealed that most early-type 
galaxies with $5\times10^9 < M_{\rm dyn}/M_{\odot} < 2\times10^{11}$ are rotating
fast within their effective half-light radii ($R_{\rm e}$).  Given their definition of
$M_{\rm dyn}$ as twice the dynamical mass within 1$R_{\rm e}$, and the low
fraction of dark matter within 1$R_{\rm e}$ for massive early-type galaxies
(e.g., Cappellari et al.\ 2013), the 
above observation also holds true for the stellar mass range  
$10^{10} < M_*/M_{\odot} < 2\times10^{11}$. 
While kinematical maps also now readily betray the presence of 
{\it intermediate-scale} disks (see also Figure~9 from Krajnovi\'c et al.\ 2013), 
care is still required when modeling and 
interpreting the host galaxy's distribution of stellar light and assigning it
to physical components such as disks, spheroids, etc. (e.g.\ Savorgnan \&
Graham 2016a; Sil'chenko 2016).

In passing we briefly discuss the nice kinematic maps for NGC~1271 presented
in W2015.  Their Figure~6 includes a velocity map over the inner 
$\sim$12x24$\arcsec$ in which the contribution from the disk
can be seen to peak along the major-axis at $\sim$8$\arcsec$ and then decline
with radius, mimicking the behavior of the ellipticity profile seen in
Figure~\ref{fig:1D}.  In addition, their absorption line map of the $h_3$ 
asymmetric deviations of the line of-sight velocity distribution from a Gaussian reveals a disk
whose dominance reaches $r_{major} \approx 15\arcsec$,  mimicking the
behavior of the $B_4$ profile seen in Figure~\ref{fig:1D}. 
The dynamical model shown in their Figure~6 can also be seen to rotate with a
greater velocity at large radii than is supported by the data.  
Figure~9 in W2015 (based around their 11-component luminous mass model) 
shows the gradual returning importance of the bulge beyond 
$\sim$10--16$\arcsec$, as is also seen in our decomposition
(Figure~\ref{fig:1D}).  We expect that more extended, major-axis kinematic data 
(beyond 24$\arcsec$) will solidify the embedded/truncated nature of the disk
in NGC~1271 and support our two-component model for the bulk of this galaxy's
stellar flux. 
Within the isophote having a major-axis equal to 24'', our model has some 77\% of
the flux in the spheroidal component and 23\% in the flattened disk.  
This is at odds with W2015 who reported that 
``the bulge component accounts for 12\% of the mass, whereas the 
rotating components total 75\% of the mass within the radial 
extent [$r_{major} \approx 24\arcsec$] of the kinematic measurements''.

Graham (2013) suggested that many of the compact
spheroids at $z\sim 2\pm0.5$ (e.g.\ Weinzirl et al.\ 2011; Damjanov et al.\ 2014;
van Dokkum et al.\ 2015, and references therein) may not have grown in size --- as commonly
thought --- because compact spheroids are still around us today
(Figure~\ref{fig:ab}), with NGC~1271 another such example. 
Indeed, the mass, half-light radius, velocity dispersion and rotation of
NGC~1271 match reasonably well with the recently quenched galaxy RG1M0150 observed 
at $z = 2.636$ by Newman et al.\ (2015). 
It is just that most local, compact massive spheroids are now connected with stellar 
 disks (see the simulations by, and Figure~2 in, 
Wellons et al.\ 2016 which supports this growth path). 
Rather than transforming these compact spheroids through minor dry mergers,
Graham (2013) speculated that disk building might have been in operation. 

This represents a fundamentally different formation path and evolutionary
history: the former is thought to build elliptical galaxies, while the latter builds both S0
galaxies\footnote{See Bois et al.\ (2011) and Querejeta et
  al.\ (2015) for an alternative pathway to build some lenticular galaxies
  from spiral galaxies.} and ES galaxies.  

This is a simple and elegant solution which has since been supported by 
the results in de la Rosa et al.\ (2016) and Margalef-Bentabol et 
al.\ (2016).  Moreover, it matches with evidence for nascent disks at high-$z$ 
(e.g.\ Longhetti et al.\ 2007; van der Wel et al.\ 2011), 
and with new 870 $\mu$m continuum emission observations of hot dusty galaxies at $z\sim2$ 
which reveal the presence of compact massive bulges in pre-existing 
large-scale disks (Tadaki et al.\ 2016; Barro et al.\ 2016).  If these disks subsequently grow in
stellar mass, and thereby effectively increase their {\it galaxy's} half light radius, they will
then be matched with observations of local lenticular galaxies.  
If, instead,  (3D envelope)-building minor-mergers was the common mechanism for growing 
the size of (many of) the compact massive galaxies seen at $z\sim2$, then most of the massive early-type
galaxies in the local universe (not to be confused with the most massive
galaxies in the local universe) should be pressure-supported elliptical 
galaxies. They are not. 
Furthermore, dry minor mergers are at odds with the $z\approx 1.2$ 
galaxy size/age results of Williams et 
al.\ (2016), and minor mergers can destroy stellar disks  
according to Khochfar et al.\ (2011).  Moreover, too much merging will introduce
scatter not seen in the galaxy scaling relations (Nipoti et al.\ 2009). 
Although, it is noted that satellite galaxies display a tendency to reside in a plane (Kroupa et
al.\ 2005; Welker et al.\ 2015), possibly favoring disk formation.

While large, massive elliptical galaxies are known to exist at $z\sim$2--3
(e.g.\ Mancini et al.\ 2010; Newman et al.\ 2010; Saracco et al.\ 2010), 
and major mergers will build more by today (e.g.\ NGC 5557, Duc et al.\ 2011) ---
with the super-massive black holes from the progenitor galaxies scouring out a
partially depleted core when it is a dry merger event (e.g.\ Gualandris \&
Merritt 2012, and references therein) --- it is now recognized
that most nearby early-type galaxies with
$10^{10}<M_*/M_{\odot}<2\times10^{11}$  are comprised of a compact spheroidal 
component plus a stellar disk 
(e.g.\ Emsellem et al.\ 2011; Krajnovi\'c et al.\ 2013). 
This suggests that disk growth, rather than messy mergers, may be how many of the compact
massive galaxies at $z\sim$2--3 have grown in size. 
In this scheme, rather than
staying on the `red-sequence' since $z\sim$2--3, gas disks are accreted around the
compact, massive high-$z$ galaxies (e.g. Birnboim \& Dekel 2003; 
Kere{\v s} et al.\ 2005, 2009; Dekel et al.\ 2009), 
stars then form in these disks and the galaxies move to the 
`blue cloud' and likely partake in high star formation rates before the disks
then rapidly reduce much of their star formation by $z\sim 1\pm0.2$ (e.g.\ Madau
et al.\ 1998; Dickinson et al.\ 2003; P\'erez-Gonz\'alez et al.\ 2005) to
produce the more massive ES and S0 galaxies around
us today.  Due to `downsizing' (Noeske et al.\ 2007; P\'erez-Gonz\'alez et al.\ 2008a,b), 
the tail-end of disk growth in less massive early-type galaxies is still
  occurring in some early-type galaxies today (e.g.\ Lemonias et al.\ 2011;
  Moffett et al.\ 2012; Alatalo et al.\ 2013; Bayet et al.\ 2013; Ger\'eb et al.\ 2016; 
GDS15, see their section~4.1 and references therein).
This scenario of disc growth 
is not just a solution to the fate of the compact massive $z\sim$2--3
galaxies, but simultaneously answers another question which oddly has not been
asked: Where, in the $z\sim$2--3 universe, are the precursors of the old,
compact massive spheroids seen in local ellicular\footnote{While working on
  this project we informally used this term to
  denote the `disk elliptical' ES galaxies intermittent between elliptical 
and lenticular galaxies.} (ES), 
lenticular (S0), and massive spiral (Sp) galaxies?

\subsection{Placing galaxies with intermediate scale disks in a familiar
  classification diagram}

The existence of intermediate-scale disks led to suggestions of a likely
continuum of disk sizes in early-type galaxies (see Simien \& de Vaucouleurs
1986; Capaccioli et al.\ 1988; Bender 1989; Simien \& Michard 1990; Scorza
1993; Scorza \& van den Bosch 1998).  The presence of these intermediate-scale
disks blurs the distinction between elliptical (E) and lenticular (S0)
galaxies, or slow-rotators (SR) versus fast-rotators (FR), see Cappellari et
al.\ (2011b, their Figures~1 and 2). That is, there is a need for more than
two bins (i.e.\ nuclear vs.\ large-scale disk, E vs.\ S0, SR vs.\ FR), and a
need for a continuum in (Hubble-Jeans)\footnote{As noted by van den Bergh
  (1997) and Sandage (2005), it was Sir James Jeans (1928) who introduced the
  (tuning fork)-shaped diagram that encapsulated Hubble's (1926)
  elliptical-spiral sequence --- itself motivated by the earlier work of Jeans
  (1919).}-like classification schemes.

It has been argued that the classification scheme for ``elliptical''
galaxies should not be their apparent ellipticity, set by their
observed axis ratio (originally
E0--E7, see Hubble 1936 and Sandage 1961; then E0--E4 when it was realised
that the E5-E7 galaxies are S0s, see Liller
1966 and also Gorbachev 1970), because this depends on our viewing
angle rather than being intrinsic to the galaxy. Obviously when
dealing with intermediate-scale disks, the disky shape of the
isophotes is a function of both the disk's inclination {\it and} the
isophotal radius, making 
this single isophotal shape parameter problematic and particularly
inappropriate for the ES class.  Section 4 of Kormendy \& Bender
(1996) points out additional complications in regard to this. 

Early-type galaxies may be better quantified by their
spheroid-to-total flux ratio, with a continuum from pure elliptical
galaxies to disk-dominated lenticular galaxies (e.g.\ Capaccioli et
al.\ 1988, and references therein). This spheroid-to-total flux (or
better mass) ratio is widely recognized as an important quantity, and
its broad range observed in the lenticular galaxies led Cappellari et
al.\ (2011b) to present the ``Hubble comb''.   This built on van den
Bergh's (1976) diagram in which both late-type galaxies (i.e.\ spiral
galaxies) and early-type galaxies form separate prongs of an expanded
Hubble-Jeans tuning fork, with an additional prong for disk galaxies
hosting anaemic spiral patterns.  Cappellari et al.\ (2011b, see also
Kormendy \& Bender 2012) presented a scheme in which the morphological
type (a, b, c, etc.) changed if the spheroid-to-total ratio
changed. {\it However}, the primary criteria used to assign the spiral galaxies'
morphological type (a, b, c, etc.) over the last half century (Sandage 1961) has been the nature of
the spiral arms --- the extent to which they are unwound, pitch angle,
and the degree of resolution --- rather than the spheroid-to-total
ratio. Indeed, in the Hubble Atlas of Galaxies, Sandage (1961) notes
that Sa type galaxies exist with both small and large bulges, see also
the ``Morphological Catalogue of Galaxies'' (MCG) 
from Vorontsov-Vel'Yaminov \& Arkhipova (1962) and
Vorontsov-Vel'Yaminov \& Noskova (1973).  

The early-type galaxies, as with the Sa type galaxies, and the Sb type
galaxies, and the Scd type galaxies, etc.\ all have a range of bulge-to-total
mass ratios.  Despite recent claims to the contrary, the fast rotating
early-type galaxies do not span the same full range of bulge fractions as
spiral galaxies. Indeed, there are bulgeless spiral galaxies but there are not
bulgeless early-type galaxies.  Morphological types have been meticulously
assigned to thousands of spiral galaxies, but because this was not primarily
based on the bulge-to-total ratio, it obviously can not now be used to
represent the bulge-to-total ratio in a revised Hubble-Jeans tuning fork
diagram.  To do so would be a mistake.  One could however ignore this
morphological type information and instead use bulge-to-total ratio as one
axis, with additional markers indicating if the galaxy has a strong or weak or
no bar, or has a strong or anaemic or no spiral pattern.

\begin{figure}
\includegraphics[trim=1cm 1cm 1cm 1cm, width=\columnwidth]{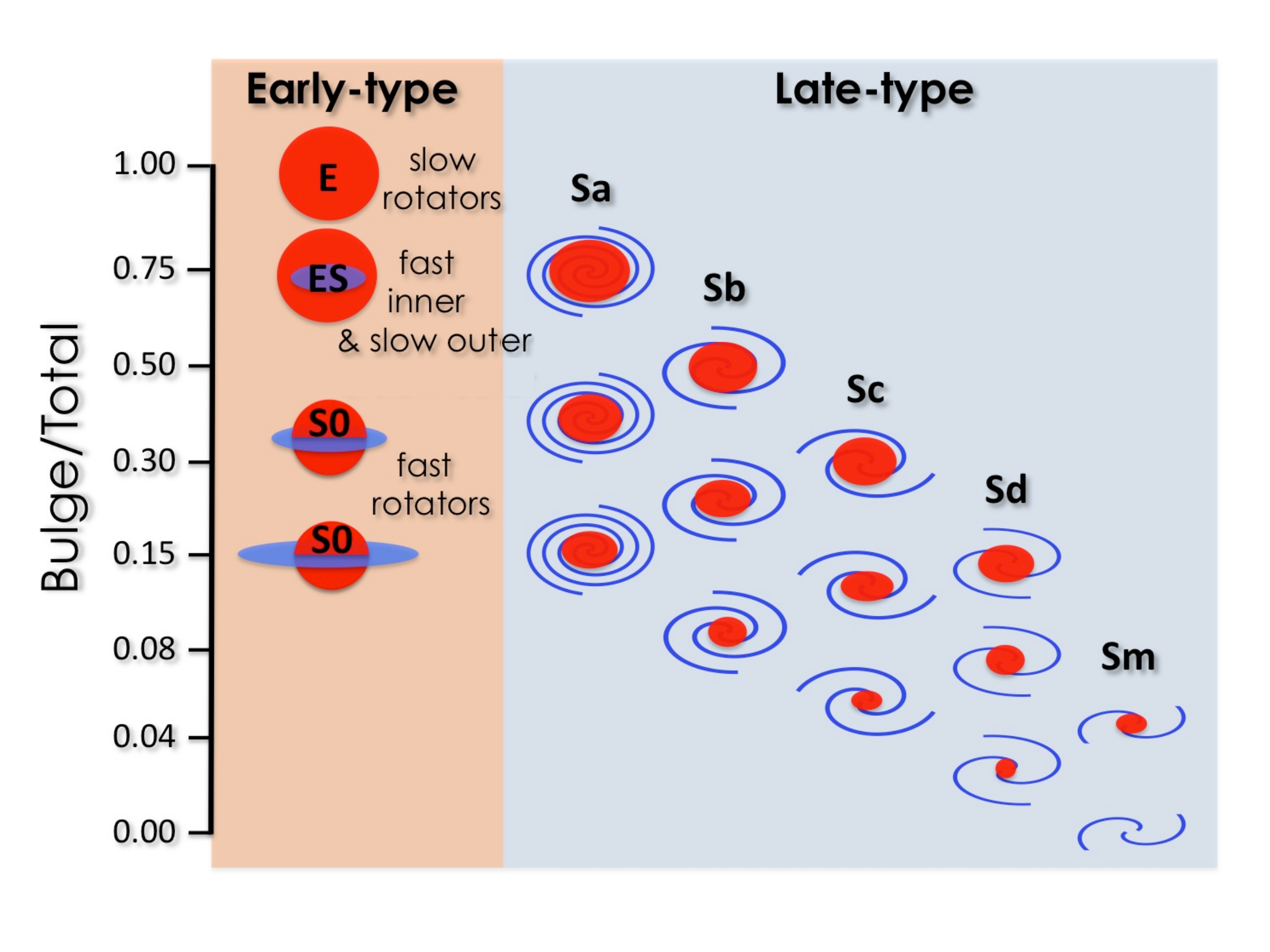}
\caption{
Grid showing the location of {\it ellicular} (ES) galaxies, 
intermediate between 
elliptical (E) and lenticular (S0) galaxies, relative to the spiral galaxies.
The horizontal axis shows the morphological type, primarily dictated by the
nature of the spiral arms.}
\label{fig:Hubble}
\end{figure}

The grid seen in Figure~\ref{fig:Hubble} and discussed in Graham
(2014) is a slight variant of that used by Freeman (1970, his Figure~9), Boroson (1981), Kent
(1985), Kodaira et al.\  (1986), Simien \& de Vaucouleurs (1986) and
others, and allows for the varying spheroid-to-total ratio among each
galaxy type (a, b, c, etc.).  This older classification scheme provides a
complementary view of the galaxies; it captures both the 
morphological type information along one axis and the spheroid-to-total ratio
along the other. 
Importantly, Figure~\ref{fig:Hubble} shows where
the early-type galaxies with intermediate-scale disks (ES) reside
relative to the other Hubble galaxy types in this familiar diagram
which has tended not to include the E and ES galaxies to date.  Rather
than using the extended classification scheme for early-type galaxies
(E - E$^+$ - L$^-$ - L$^o$ - L$^+$ - S0, corresponding to the
numerical T-type ranging from $-$5 to 0, e.g., de Vaucouleurs et
al.\ 1964, Heidmann et al.\ 1972), they are grouped together in the single
left-hand column of Figure~\ref{fig:Hubble} instead of extended
horizontally to the left (e.g.\ Andreon \& Davoust 1997, their section~5). 
While, collectively, spiral galaxies
typically display a range of spheroid-to-total flux ratios from 0 to
3/4 (e.g. Graham \& Worley 2008, and references therein), lenticular
galaxies commonly display a range from 0.1 to 3/4 (Laurikainen et
al.\ 2010), and the ES class have larger ratios.

\section{Conclusions}\label{Sec_Conc}

The black hole mass predicted by W2015 was lower than they expected 
because the embedded, intermediate-scale disk in
NGC~1271 (evident in Figure~\ref{fig:image} and the lower panels of Figure~\ref{fig:1D}) 
led to some confusion as to what is the `spheroidal' component of this 
galaxy.  Some early-type galaxies have not managed to grow their disks into
the more easily identified large-scale disks seen in many lenticular
galaxies.  When this happens, the spheroidal component of these ES galaxies remains
larger than their disk component, and thus the spheroid not only appears as a central
``bulge'' but it also dominates the light at large radii.  Having modeled the
distribution of light in NGC~1271 with an embedded intermediate-scale disk 
(Figure~\ref{fig:1D-B} and the upper panels of Figure~\ref{fig:1D},
see also Figure~\ref{fig:2D}), we report that the 
full spheroid stellar mass in NGC~1271 is $9.0\times 10^{10}~M_{\odot}$, 
greater than that adopted by W2015 ($5.4\times10^{10}~M_{\odot}$).
We additionally report that 
this spheroid is consistent in the size-mass diagram with 
other (relic) spheroids found in massive early-type disk galaxies
(Figure~\ref{fig:ab}).  

Much of the thrust of the W2015 paper arose from the suspected order of magnitude
offset of NGC~1271 from a near-linear $M_{\rm bh}$--$M_{\rm sph,*}$ relation
in the $M_{\rm bh}$--$M_{\rm sph,*}$ diagram. However we find that NGC~1271 is
not an unusual outlier in this diagram (Figure~\ref{fig:M-M}). 
The black hole mass in NGC~1271 is only
1.6-sigma above the near-linear 
$M_{\rm bh}$--$M_{\rm sph,*}$ relation for early-type galaxies (Savorgnan et
al.\ 2016). 
Therefore, the idea that NGC~1271 may represent a galaxy whose black hole and
host spheroid co-evolved in accordance with the steep near-quadratic $M_{\rm
  bh}$--$M_{\rm sph,*}$ relation to particularly high masses --- as discussed
in Graham \& Scott (2015) --- and resulting in a {\it significant} departure from
other massive spheroids, is not supported by the data (see Figure~\ref{fig:M-M}).

\acknowledgments

This research was supported under the Australian Research Council 
funding scheme (FT110100263). 
G.A.D.S.\ warmly thanks Chien Peng for useful discussions. 
This research has made use of the NASA/IPAC Extragalactic Database (NED). 
Based on observations made with the NASA/ESA {\it Hubble Space Telescope}, obtained
through program GO-13050.

\end{document}